\documentclass[12pt,aps,showpacs,preprintnumbers,amsmath,amssymb,amsfonts,superscriptaddress,openany,nofootinbib]{revtex4-1}
\usepackage[utf8]{inputenc}
\usepackage{amsfonts}
\usepackage{amsbsy}
\usepackage{url}
\usepackage{enumerate}
\usepackage[justification=raggedright]{caption}
\usepackage{bbm}
\usepackage{mathdots}
\usepackage{graphicx}
\usepackage{graphicx,color}
\usepackage{amssymb,amsmath}
\usepackage{slashed}
\usepackage{hyperref}
\usepackage{braket}
\usepackage{simplewick}
\usepackage[title]{appendix}
\usepackage{caption}
\usepackage{ascmac}
\usepackage{latexsym}
\usepackage{pifont}          
\usepackage{bm}
\newcommand{\be}{\begin{eqnarray}}
\newcommand{\ee}{\end{eqnarray}}

\newcommand{\p}{\partial}

\begin{document}

\preprint{YGHP-18-09}

\title{
 Massless bosons on domain walls\\
-- Jackiw-Rebbi-like mechanism for bosonic fields --}

\author{Masato Arai}
\affiliation{Faculty of Science, Yamagata University,
Kojirakawa-machi 1-4-12, Yamagata, Yamagata 990-8560, Japan}

\author{Filip Blaschke}
\affiliation{Institute of Physics and Research Centre of 
Theoretical Physics and Astrophysics,
Faculty of Philosophy and Science, Silesian University in Opava,
Bezru\v{c}ovo n\'am. 1150/13, 746~01 Opava, Czech Republic}
\affiliation{Institute of Experimental and Applied Physics, 
Czech Technical University in Prague,
Horsk\'a 3a/22, 128 00 Praha 2, Czech Republic}

\author{Minoru Eto}
\affiliation{Faculty of Science, Yamagata University,
Kojirakawa-machi 1-4-12, Yamagata, Yamagata 990-8560, Japan}
\affiliation{Research and Educational Center for Natural 
Sciences, Keio University,
Hiyoshi 4-1-1, Yokohama, Kanagawa 223-8521, Japan}

\author{Norisuke Sakai}
\affiliation{Research and Educational Center for Natural 
Sciences, Keio University,
Hiyoshi 4-1-1, Yokohama, Kanagawa 223-8521, Japan}
\affiliation{iTHEMS, RIKEN,
2-1 Hirasawa, Wako, Saitama 351-0198, Japan }

\begin{abstract}
\ \\
It is important to obtain (nearly) massless localized modes 
for the low-energy four-dimensional effective field theory in 
the brane-world scenario. 
We propose a mechanism for bosonic zero modes using the 
field-dependent kinetic function in the classical field theory 
set-up. 
As a particularly simple case, we consider a domain wall 
in five dimensions, and show that massless states 
for scalar (0-form), vector (1-form), and tensor (2-form) fields 
appear on a domain wall, which may be called topological because 
of robustness of their existence (insensitive to continuous 
deformations of parameters). 
The spin of localized massless bosons is selected by the shape 
of the nonlinear kinetic function, analogously to the chirality 
selection of fermion by the well-known Jackiw-Rebbi mechanism. 
Several explicitly solvable examples are given. 
We consider not only (anti)BPS domain walls in non-compact extra 
dimension but also non-BPS domain walls in compact extra dimension.
\end{abstract}

\maketitle

%%%%%%%%%%%%%%%%% I N T R O D U C T I O N %%%%%%%%%%%%%%%%%%

\section{Introduction}

A long time ago, Jackiw and Rebbi showed that massless fermions are trapped 
by a topological soliton, namely a domain wall \cite{Jackiw:1975fn}.
As it turns out, this property is robust since it depends on 
topological aspects of a given theory alone and it is otherwise 
insensitive to the details.
This idea has become ubiquitous within a vast area of modern physics.
Let us give several examples. 
Topological kinks in polyacetylene are described by Su, Schrieffer, 
and Heeger \cite{Su:1979ua}, and quantized solitons of the 
one-dimensional Neel state are studied by Haldane \cite{Haldane:1983ru}.
Rubakov and Shaposhnikov \cite{Rubakov:1983bb} studied the 
possibility that our (3+1)-dimensional universe is embedded 
in higher dimensions, which is an early proposal of the so-called 
brane-world scenario 
\cite{ArkaniHamed:1998rs,Antoniadis:1998ig,Randall:1999ee,Randall:1999vf}.
The Jackiw-Rebbi mechanism naturally provides massless chiral 
fermions (leptons and quarks) on a domain wall (a $3$-brane) in 
five dimensions. 
The left- or right-handed chirality is selected by 
the profile of the domain wall (kink) background solution. 
The mechanism has also been used to treat chiral fermions in lattice QCD,
the so-called domain wall fermion, in 
Refs.~\cite{Kaplan:1992bt,Shamir:1993zy,Furman:1994ky}.
Furthermore, there is an intimate connection between the 
Jackiw-Rebbi mechanism and a topological phase of matter which is one of
the highlights in the last decade. There, an interplay between topology and
massless edge (surface) modes has revealed new, rich properties of 
matter \cite{Hasan:2010xy,Qi:2011zya}.

These massless modes on edges are all fermionic states.
 Thus, we are lead to a natural question:
Do massless bosons, especially gauge bosons, also robustly 
appear on domain walls (edges)? In this paper, 
we answer this question in the affirmative.
 
We arrived at this question not under the necessity of application 
to some materials in condensed matter.
Rather, we have encountered it in our recent studies on quite 
different topic, the dynamical construction of brane-world 
scenario by topological solitons \cite{Arai:2012cx,Arai:2013mwa,
Arai:2017ntb,Arai:2017lfv,Arai:2018rwf,Arai:2018uoy}.
A necessary condition common to most brane-world models is 
that all Standard Model particles, except for four-dimensional 
gravitons, must be localized on the 3-brane\footnote{
We are assuming the extra dimensions to be noncompact 
or large.
}. 
Namely, fermions, scalar and vector bosons must be localized 
on the 3-brane. 
It is desirable for a localization mechanism not to depend 
on details of the model.
The Jackiw-Rebbi mechanism is indeed a prime example of such 
a mechanism\footnote{
\label{FN:fractional_fermion}
In $1+1$ dimensions, a domain wall coupled to fermions 
may be considered as degenerate fermionic-soliton states with 
fractional fermion numbers \cite{Jackiw:1975fn}. 
On the other hand, we interpret a localized fermion on a domain 
wall in higher dimensions, say $4+1$ dimensions, as an elementary 
fermionic particles such as quarks and leptons confined inside 
the domain wall \cite{Rubakov:1983bb}.
}, providing chiral fermions on a domain wall (3-brane) \cite{Rubakov:1983bb}.
How about bosons?
The Standard Model also has bosonic fields: the Higgs field and $SU(3)\times SU(2) \times U(1)$ gauge bosons.
Unlike fermions, however, a robust localization mechanism for bosons, especially non-Abelian Yang-Mills fields, is not widely agreed on.
There were many works so far 
\cite{Dvali:2000rx, Kehagias:2000au, Dubovsky:2001pe, 
Ghoroku:2001zu,Akhmedov:2001ny, Kogan:2001wp, Abe:2002rj, 
Laine:2002rh, Maru:2003mx, Batell:2006dp, Guerrero:2009ac, 
Cruz:2010zz, Chumbes:2011zt, Germani:2011cv, Delsate:2011aa, 
Cruz:2012kd, Herrera-Aguilar:2014oua, Zhao:2014gka, Vaquera-Araujo:2014tia,
Alencar:2014moa,Alencar:2015awa,Alencar:2015oka,Alencar:2015rtc,Alencar:2017dqb,Zhao:2017epp}. 
Among them, one of the most popular idea relies on strongly coupled dynamics: 
a domain wall in confining vacua. A concrete model in four spacetime dimensions was 
explicitly proposed \cite{Dvali:1996xe}. 
Due to the so-called dual Meisner effect, (chromo)electric 
field cannot invade the bulk, so that massless gauge fields 
are confined inside the wall.
This mechanism is clearly  independent of the details. 
However, since it is based on strong coupling dynamics which is not very well understood in four let alone five dimensions,
it is very hard to quantitatively deal with any physics related to massless four-dimensional gauge fields.  
Therefore, in practice the confinement in higher dimensions was simply assumed to take place,
see for example Refs.~\cite{Libanov:2000uf,Frere:2000dc,Frere:2001ug,Frere:2003yv,Volkas,Volkas2,Callen:2010mx}.

Alternatively, a phenomenological model with a field-dependent kinetic term for gauge fields was 
considered in six spacetime dimensions \cite{ArkaniHamed:1998rs}. 
One does not need to assume confinement in higher dimensions.
Rather, it can be thought of as  
an effective description of confinement in terms of classical 
fields \cite{Kogut:1974sn,Friedberg:1976eg,Friedberg:1977xf,
Friedberg:1978sc,Fukuda:1977wj,Fukuda:2008mz,Fukuda:2009zz}.
Hence, one can quantitatively study phenomena involving the massless four-dimensional gauge fields.
A supersymmetric model has been constructed in 
five spacetime dimensions \cite{Ohta:2010fu}, and further 
developments into unified theories beyond the Standard 
Model followed \cite{Arai:2012cx,Arai:2013mwa,Arai:2017ntb,Arai:2017lfv,
Arai:2018rwf,Arai:2018uoy}, see also \cite{Okada:2017omx,Okada:2018von}.
A detailed study of localization by the field-dependent gauge kinetic terms was done earlier in \cite{Dubovsky:2001pe}, and
another study for nonsupersymmetric model with/without gravity 
was developed in \cite{Chumbes:2011zt}, see also a recent review paper \cite{Liu:2017gcn}.
 
In this paper, we will reanalyze the localization of 
massless gauge fields on a domain wall via the field-dependent 
gauge kinetic term from a different viewpoint where we 
 do not need the speculative connection between it and confinement. 
Instead, we find a common mathematical structure 
and a mapping between our localization mechanism of gauge fields 
and the Jackiw-Rebbi mechanism for fermions. 
We call this underlying mathematical structure for bosons 
as Jackiw-Rebbi-like mechanism for bosons.
As we will show explicitly, the presence of massless gauge fields 
on a domain wall relies only on boundary conditions. Thus, it 
is topological in the sense that it does not depend on precise 
form of the Lagrangian. Once we recognize the massless gauge 
fields as topological, we will show that the 
Jackiw-Rebbi-like mechanism 
for bosons works not only vector (1-form) fields 
but also for scalar (0-form) and tensor (2-form) fields.
Since there is no obvious reason for massless 0- and 2-form 
tensor fields to be related to confinement, the 
Jackiw-Rebbi-like mechanism for bosons is a nice and concrete 
explanation alternative to the confinment.
We will work on domain walls in 5 dimensions in this work. 
Similarly to the selection of
chirality of four-dimensional fermion 
by the wall, we will show the Jackiw-Rebbi-like mechanism selects the 
spin of localized massless bosons: It selects between 
four-dimensional vector 
or scalar (tensor or vector) in the case of 
five-dimensional vector (tensor) 
bosonic fields.

Here, let us make distinctions between this paper and  
previous works clear.
First of all,  this work  presents a different  point of view that
the Jackiw-Rebbi-like mechanism plays a main role 
for the localization. Admittedly, there is a partial overlap 
between the models we study 
in Sec.~\ref{sec:cal} and those in Ref.~\cite{Chumbes:2011zt}.
However, treatment of extra components of bosonic fields 
(components perpendicular to the domain wall; 
$A_y$ for vector fields and $\theta_{\mu y}$ for tensor fields) are
clearly different. We do not take the axial gauge of $A_y=0$ 
(We will explicitly show that the axial gauge is inadequate to 
discover massless modes).  
This is especially important if we consider a pair of a wall 
and an anti-wall in a compact extra dimension since additional 
physical massless bosons arise from $A_y$ and $\theta_{\mu y}$
as we will show in Sec.~\ref{sec:non-BPS}.

The organization of the paper is as follows.
We briefly describe well-known facts about domain walls in Sec.~\ref{sec:DW}.
Topological edge states are explained in Sec.~\ref{sec:zero_wall}.
In the first subsection we review the Jackiw-Rebbi mechanism for 
fermions and the rest is devoted for scalar, vector, and tensor bosonic fields.
We  provide several  explicit models in Sec.~\ref{sec:example}.
Only in Sec.~\ref{sec:non-BPS}, we consider a pair of a  wall and an anti-wall
 with a compact extra dimension. 
 Phenomenological implications are also discussed.

%%%%%%%%%%%%%%%%%%%%%%%%%%%%%%%%%%%

\section{Domain walls: A brief review}
\label{sec:DW}

Let us consider a scalar model in non-compact flat 
five-dimensional spacetime\footnote{We will consider five 
dimensions in order to provide a brane-world 
model by a dynamical compactification \cite{Dvali:1996bg}. 
However, in general, one can consider more (or less) dimensions without 
 significant changes. 
} 
 ($D=5$)
\be
{\cal L}_{\rm DW} = \p_MT \p^MT  - W_T^2\,,\quad (M=0,1,2,3,4)\,,
\label{eq:L}
\ee
where we have expressed, for later convenience, a scalar potential 
$V(T) = W_T(T)^2$  in terms of a ``superpotential'' 
$W(T)$ which is an arbitrary function of a real scalar field $T$. 
Hereafter we use the notation such as 
\be
W_T = \frac{dW}{dT},\qquad W_{TT} = \frac{d^2W}{dT^2}.
\ee

We assume that there exist multiple discrete vacua satisfying $W_T = 0$.
Let $T=T(y)$ be a domain wall solution which interpolates adjacent 
vacua at $y=\pm \infty$
($y$ stands for one of the spatial coordinates). 
 The static equation of motion reads
\be
-T'' + W_T W_{TT} = 0\,,
\label{eq:EOM}
\ee
where the prime denotes a derivative in terms of $y$.
Let us investigate  the mass spectrum by perturbing $T$ about the
background domain wall solution as $T(y)  \to T(y) + \tau(x^\mu,y)$ 
with $\tau$ being a small fluctuation of the scalar field. 
The linearized equation of motion is found as
\be
\left(\square - \p_y^2 + W_{TT}^2 + W_T W_{TTT}\right) \tau = 0,
\label{eq:LEOM_T}
\ee
where $W_{T}, W_{TT}$, and $W_{TTT}$ should be understood as those 
evaluated at the domain wall solution $T=T(y)$.
Hence, the mass spectrum is determined by solving the 
eigenvalue problem in one dimension with the $n$-th eigenfunction 
$g_n$ corresponding to the mass squared eigenvalue $m_n^2$ 
\be
\left(- \p_y^2 + W_{TT}^2 + W_T W_{TTT}\right) g_n = m_n^2 g_n.
\label{eq:Sch_scalar}
\ee
Irrespective of the details of the superpotential $W$, there always 
exists a normalizable zero mode. 
To see this, let us differentiate Eq.~(\ref{eq:EOM}) once by $y$
\be
\left(-\p_y^2 + W_{TT}^2 + W_TW_{TTT}\right)T' = 0.
\ee
Thus, we find a solution with zero eigenvalue (apart from the 
normalization constant) 
\be
g_0 = T'.
\ee
The presence of  this normalizable\footnote{Since we are interested in finite tension walls it follows that the zero mode is normalizable.} zero mode is robust, because 
it is nothing but the Nambu-Goldstone zero mode 
associated with the spontaneously broken
translational symmetry.

Stability of the domain wall background is ensured by topology. 
When a static configuration $T$ is a function of $y$, 
we can derive the well-known Bogomol'nyi completion form for the 
energy density ${\cal E}$ as 
\be
{\cal E} = T'{}^2 + W_T^2 = \left(T' \mp W_T\right)^2 
\pm 2 T' W_T \ge 
\pm 2W' 
\,.
\label{eq:bogomolnyi}
\ee
This Bogomol'nyi inequality is useful by choosing the 
upper (lower) sign for $W'>0$ ($W'<0$). It is saturated by solutions 
of the so-called BPS equation 
\be
T' = \pm W_T.
\label{eq:BPS}
\ee
We call the upper sign the BPS while the lower sign the antiBPS.\footnote{
The BPS solution often has the underlying supersymmetry. 
Namely the system allowing the BPS solution can usually be 
embedded into a supersymmetric theory and the BPS solution 
preserves a part of supersymmetry. 
}
Tension of the domain wall is finite since we have assumed 
a boundary condition with $T' = \pm W_T \to 0$ 
as $|y| \to \infty$.
It is straightforward to verify that any solution of the BPS 
equation solves the full EOM (\ref{eq:EOM}).
Tension of the BPS domain wall reads 
\be
\sigma = \int^\infty_{-\infty}dy\, {\cal E} = 
2 \left|W\left(T(+\infty)\right) - W\left(T(-\infty)\right)\right|.
\ee
This is a topological quantity.  To see this, let us define a 
conserved current by\footnote{
We temporarily disregard the Lorentz invariance in four-dimensional 
world volume of the domain wall by treating the time direction 
$x^0$ separately from spatial directions $x^1, x^2, x^3$.} 
\be
j^\alpha = \epsilon^{\alpha\beta}\p_\beta W(T),\qquad(\alpha,
\beta = 0,y).
\ee
Then the topological charge $q$ reads
\be
q = \int^\infty_{-\infty} dy\, j^0 = \int^\infty_{-\infty} dy\, \p_y W(T) 
=  W\left(T(+\infty)\right) - W\left(T(-\infty)\right).
\ee
After appropriately normalized, we find that the (anti)BPS 
domain wall has the topological charge $(-)1$.

If the background configuration is a BPS or an antiBPS solution 
rather than a general solution of field equation in 
Eq.~(\ref{eq:EOM}), we can obtain more precise informations 
as follows.  
Using the BPS equation $T' = W_T$, 
the eigenvalue equation (\ref{eq:Sch_scalar}) can be rewritten as
\be
\text{BPS}:&& Q^\dagger Q g_n = m_n^2 g_n\,,
\label{eq:spectrum_T_BPS}
\ee
where we have introduced 1st order differential operators
\be
Q = - \p_y + W_{TT}(T(y)),\qquad
Q^\dagger = \p_y + W_{TT}(T(y)).
\label{Eq:Q}
\ee
Similarly, for the antiBPS solution ($T' = - W_T$), 
the eigenvalue equation can be rewritten as 
\be
\text{antiBPS}:&& Q Q^\dagger g_n = m_n^2 g_n\,.
\label{eq:spectrum_T_antiBPS}
\ee
The Hamiltonians $Q^\dagger Q$ and $QQ^\dagger$ are semi-positive 
definite, so there are no tachyonic instabilities.
It is interesting to note that the above system of 
equations constitutes a supersymmetric quantum 
mechanics \cite{Witten:1981nf} (SQM). 
The SQM superpotential ${\cal X}(y)$ is defined as 
\be
Q = -\p_y + {\cal X}'. 
\ee
In this case of scalar field $T$ for the BPS domain wall, the SQM 
superpotential ${\cal X}$ is related to the ``superpotential'' 
$W$ in the $D=5$ field theory Lagrangian (\ref{eq:L}) as 
\be
{\cal X}(y)\big|_{\rm (anti)BPS} =  \pm\frac{1}{2}\log W_T(T(y))^2\,.
\label{eq:SQM_W}
\ee
By using the (anti)BPS equation,
the translational zero mode $g_0$ can be expressed as
\be
g_0(y)\big|_{\rm (anti)BPS} =  W_T(T(y)).
\label{eq:trans_0_BPS}
\ee
We emphasize that the SQM form is valid for the translational 
zero mode only if the domain wall satisfies the BPS equation. 

%%%%%%%%%%%%%%%%%%%%%%%%%%%%%%%%%%%
\section{Massless states on domain walls }
\label{sec:zero_wall}

%%%%%%%%%%%%%%%%%%%%%%%%%%%%%%%%%%%

\subsection{Domain wall fermions: A review on the Jackiw-Rebbi mechanism}
\label{sec:fermion}

In addition to scalar fields in  ${\cal L}_{\rm DW}$, let us consider a five-dimensional Dirac fermion $\Psi$ in the form  
\be
{\cal L}_{\rm F} = i \bar\Psi \Gamma^M\p_M \Psi 
- {\cal M}(T) \bar\Psi\Psi\,.
\label{eq:LF}
\ee
The gamma matrices in $D=5$ are related to those in $D=4$ by 
$\Gamma^\mu = \gamma^\mu$ and $\Gamma^4 = i\gamma_5$.
The field-dependent ``mass" ${\cal M}(T)$ is just a coupling function of scalar 
fields multiplying the term quadratic in fermion fields. 
It becomes a $5D$ fermion mass only when it is a constant 
and independent of any fields. 
We assume that the function ${\cal M}(T)$ is real. 
When considering the Kaluza-Klein decomposition to (infinitely 
many) $4D$ components, there is no reason for massless 
$4D$ fermions to exist with a generic ${\cal M}(T)$, except for the 
well-known Jackiw-Rebbi mechanism~\cite{Jackiw:1975fn}. 
The mechanism ensures the existence of massless 
fermions localized on a domain wall, and works in both even and odd 
dimensions.
The masslessness of the fermion resulting from the Jackiw-Rebbi 
mechanism is stable against small deformations of parameters. 
In this sense, the Jackiw-Rebbi fermion is topological.

\begin{figure}[t]
\begin{center}
\includegraphics[width=15cm]{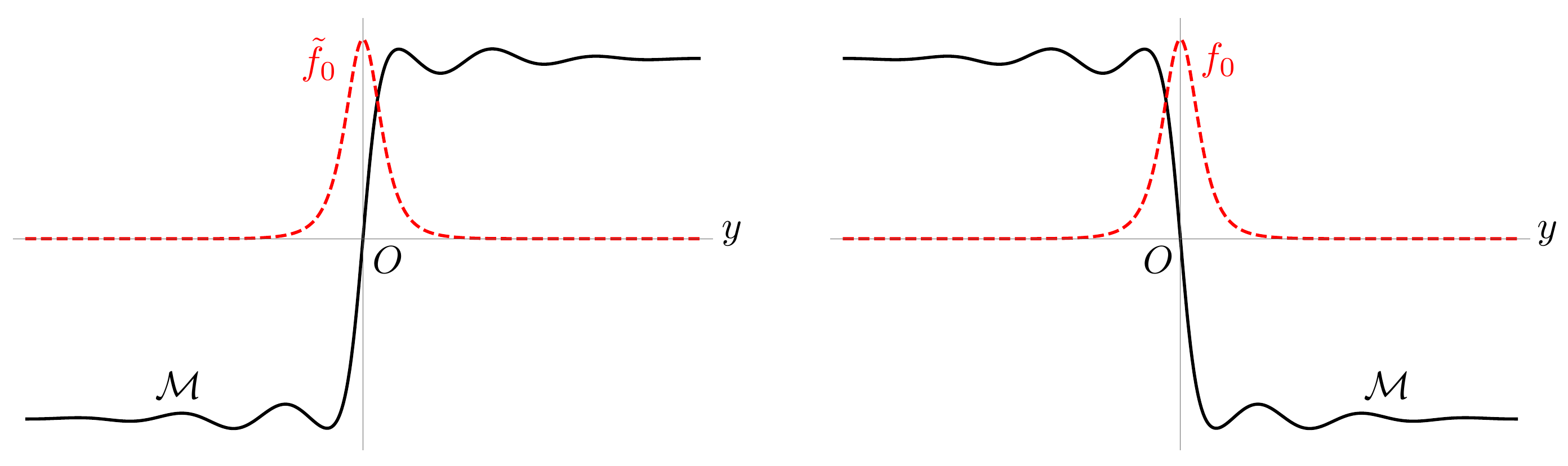}
\caption{The field-dependent ``mass" ${\cal M}(T(y))$ 
and the associated topological edge states (fermion zero modes).
The solid curves are ${\cal M} =\pm \left( \tanh y + \frac{1}{10} e^{-y^2/100}\sin y\right)$, and the broken red curves correspond
to the mode functions of the fermion zero modes.}
\label{fig:fermion_zero}
\end{center}
\end{figure}

To see how the Jackiw-Rebbi mechanism works, let us 
investigate mass spectra of the fermion around the domain 
wall background $T(y)$.\footnote{Here we do not restrict ourselves
to the (anti)BPS domain wall. The background can be non-BPS.}
We assume that asymptotic values of ${\cal M}(T(y=\pm\infty))$ 
at left and right infinity are non zero and
have opposite sign, as in the typical 
kink-like configuration, see Fig.~\ref{fig:fermion_zero}.
\be
{\cal M}(T(y=-\infty))\times {\cal M}(T(y=+\infty))<0. 
\label{eq:opp_asymp_value}
\ee
%Let us call this as the gap condition.
Linearized equations of motion for fermionic fluctuations $\Psi$ 
(using the same character $\Psi$ for the small fluctuation) 
reads
\be
i\gamma^\mu\p_\mu \Psi - \gamma_5\p_y\Psi - {\cal M} \Psi = 0\,.
\ee
Let us define a ``Hamiltonian''
\be
H_5 = - \gamma_5\p_y - {\cal M}\,.
\ee
A normalizable zero eigenstate of $H_5\left|0\right>=0$ can be
easily found by multiplying $\gamma_5$ from left
and considering eigenstates of $\gamma_5\left|\pm\right>=\pm\left|\pm\right>$ for which it holds
\be
{\cal Q} \left|-\right> = 0\,,\qquad
{\cal Q}^\dagger \left|+\right> = 0\,,
\ee
where the ${\cal Q}$ and ${\cal Q}^\dagger$ operators are defined by
\be
{\cal Q} = - \p_y + {\cal M}(y)\,,\qquad
{\cal Q}^\dagger = \p_y + {\cal M}(y)\,. \label{eq:Qf}
\ee
In the coordinate representation these states reads
\be
\left<y|-\right> \equiv f_0(y) = e^{\int^y d \lambda\, {\cal M}(T(\lambda))}\,,\qquad
\left<y|+\right> \equiv \tilde f_0(y) = e^{-\int^y d \lambda\, {\cal M}(T(\lambda))}\,,
\label{eq:fermion_zero_mode}
\ee
up to normalization constants. Since the domain wall connects 
different vacua with opposite sign for ${\cal M}(T(y=-\infty))$ 
and  ${\cal M}(T(y=+\infty))$ as in Eq.\ (\ref{eq:opp_asymp_value}), 
 ${\cal M}(T(y))$ must vanish at a finite value of $y$, usually 
around the center of the domain wall. 
When ${\cal M}(T(y))$ increasingly (decreasingly) goes across zero,
the right(left)-handed fermion is localized on the 
domain wall, see Fig.~\ref{fig:fermion_zero}.
This property does not depend on any details of the solution, and
it is the heart of the Jackiw-Rebbi model \cite{Jackiw:1975fn}.
In terms of a modern terminology, the massless fermion is 
often called the topological edge state \cite{Hasan:2010xy}.
As was mentioned in the footnote \ref{FN:fractional_fermion},
these fermions should be interpreted as four dimensional fermionic particles
confined inside the domain wall.

Let us make our statement clearer. Hereafter, we use the Jackiw-Rebbi
mechanism for fermions for the following meaning. {\it When the field
dependent ``mass'' ${\cal M}(T)$ defined in Eq.~(\ref{eq:LF})
satisfies the %gap 
condition 
%given in Eq.~
(\ref{eq:opp_asymp_value}), 
either left- or right-handed
massless fermion appears around a point where ${\cal M}$ vanishes.}
The chirality of the massless fermion is determined 
by the sign of the asymptotic value ${\cal M}(y=+\infty)$ 
: Left-handed for ${\cal M}(y=+\infty)<0$, 
and right-handed for ${\cal M}(y=+\infty)>0$. 
We also define topological particles as those massless 
particles that remain massless under continuous deformations of 
parameters, and are not explained by symmetry reasons
such as a spontaneously broken rigid symmetry.
The domain wall fermion is a typical topological particle\footnote{
For completeness, let us briefly mention here another known 
physical reason to ensure masslessness of a fermion: 
the Nambu-Goldstone (NG) fermion \cite{Fayet:1974jb,ORaifeartaigh:1975nky} 
as a result of the spontaneously broken rigid fermionic symmetry 
such as supersymmetry.
The masslessness of the NG fermion is stable against small deformations 
of parameters, protected by a symmetry reason. 
In contrast, instead of symmetry, the domain wall fermion realized by the Jackiw-Rebbi mechanism is protected 
by a topological reason.
}
which does not disappear against any continuous changes without 
violating the 
%gap 
condition given in Eq.~(\ref{eq:opp_asymp_value}).  

For later uses, let us give a complete analysis for the mass spectra. Firstly, we decompose $\Psi$ into $\Psi_L$ and $\Psi_R$ which are the 
eigenstates of $\gamma_5$ as
$\gamma_5 \Psi_L = - \Psi_L$ and $\gamma_5 \Psi_R = \Psi_R$. We find
\be
i\gamma^\mu\p_\mu \Psi_R = {\cal Q} \Psi_L,\qquad
i\gamma^\mu\p_\mu \Psi_L = {\cal Q}^\dagger \Psi_R\,.
\ee
Eliminating $\Psi_R$ ($\Psi_L$), we reach the following equations 
\be
\left(\square  + {\cal Q}^\dagger {\cal Q}\right) \Psi_L = 0,\quad
\left(\square  + {\cal Q} {\cal Q}^\dagger \right) \Psi_R = 0.
\label{eq:SQM_fermion}
\ee
Thus, the physical spectra for $\Psi_{L,R}$ are determined by 
solving the 1D eigenvalue 
problems 
\be
{\cal Q}^\dagger {\cal Q} f_n = M_n^2 f_n,\qquad
{\cal Q} {\cal Q}^\dagger \tilde f_n = M_n^2 \tilde f_n.
\label{eq:Sch_f}
\ee
We again encounter a 1D SQM problem with the superpotential 
${\cal Q}=-\partial_y+{\cal Y}'(y)$ given in (\ref{eq:Qf}),
\be
 {\cal Y}(y)=\int^y d \lambda\ {\cal M}(T(\lambda)).
\ee

We would like to
emphasize that this formula is correct regardless 
of whether the background 
solution is (anti)BPS or non-BPS. This is in contrast to the fluctuation 
of $T$ field given in Eq.~(\ref{eq:spectrum_T_BPS})
or (\ref{eq:spectrum_T_antiBPS}) which are valid only for the 
(anti)BPS background solution.
As before, the 1D Hamiltonians are semi-positive definite, so 
that there are no tachyonic modes. Furthermore, 
due to the SQM structure, 
$\Psi_L$ and $\Psi_R$ share the identical mass spectra 
except for possible zero modes, in accord with the fact that 
any modes with a nonvanishing mass consist of both chiralities 
in even dimensions.

We will now turn to massless bosons in subsequent sections.

%%%%%%%%%%%%%%%%%%%%%%%

\subsection{Domain wall scalars}
\label{eq:DWS}

Contrary to fermions, the protection mechanism for 
masslessness of scalar fields is not known\footnote{We are 
aware of the fact that supersymmetry combined with the chiral 
symmetry can protect the masslessness of the scalar particle 
accompanied by the massless fermion \cite{Witten:1981nf,
Dimopoulos:1981zb,Sakai:1981gr}. This idea has been extremely 
popular and productive, though it may be regarded as somewhat 
indirect. } except for the symmetry 
reason associated with the spontaneously broken rigid
symmetry 
with a continuous parameter, namely the Nambu-Goldstone boson. 
For example, we found in Sec.~\ref{sec:DW} a normalizable scalar 
zero mode on the domain wall background, whose existence 
is ensured by the spontaneously broken 
translational symmetry.

Guided by the Jackiw-Rebbi mechanism for fermions, one might 
be tempted to try considering a real scalar field $\Phi$ whose 
coupling function for quadratic term is given by the same field-dependent ``mass" 
${\cal M}(T)$ as 
in Eq.~(\ref{eq:LF}): 
\be
{\cal L}_{\rm S} = \frac{1}{2}\p_M\Phi\p^M\Phi - 
\frac{1}{2}{\cal M}(T)^2 \Phi^2,
\ee
in addition to the Lagrangians (\ref{eq:L}).
Since ${\cal M}(T)^2 \Phi^2$ is semi-positive definite, 
$\Phi$ remains inert as $\Phi=0$, when $T$ takes the domain 
wall configuration as a solution of the equation of motion. 
Since the 1D eigenvalue problem for the fluctuation of $\Phi$ 
on this background has a positive definite potential, ${\cal M}(T(y))^2$, it is obvious 
that there are only massive modes. 
This illustrates that the naive attempt does not work
for bosons.

We now wish to propose a mechanism for a domain wall 
scalar boson, namely a model with a massless scalar mode whose existence is insensitive to change of parameters. 
Instead of tuning a scalar potential, we turn to use a nonlinear 
kinetic term with a field-dependent kinetic function. 
Let us assume the following simple Lagrangian in addition 
to ${\cal L}_{\rm DW}$:  
\be
{\cal L}_0 = 
\beta(T)^2
\p_M\Phi\p^M\Phi.
\label{eq:L_scalar}
\ee
A field-dependent ``coupling'' $\beta(T)$ is a function of
the scalar field $T$ multiplying the term quadratic in
$\p_M \Phi$.
This form is inspired by nonlinear kinetic function 
for gauge and form fields, which are described in subsequent sections.
One can characterize 
absence of a potential for $\Phi$
as a result of a ``shift" symmetry
$\Phi \to \Phi + {\rm constant}$.
We do not consider a mixed term like $g(T)\p_M T\p^M\Phi$ in this paper, since adding it is a large deformation in the sense that it
changes the structure of Lagrangian qualitatively.
Alternatively, one can forbid it by
imposing the parity $\Phi \to -\Phi$.

Vacuum condition is $\Phi=\text{const.}$ and $W_T = 0$. As before, we assume that
there are several discrete vacua.
Then, $T$ has a nontrivial domain wall configuration whereas 
$\Phi = \text{const.}$ as a background solution. 
As for the mass spectra of fluctuations on the background 
domain wall solution, 
the linearized equation for the $T$ field is unchanged from 
Eq.~(\ref{eq:LEOM_T}).
Therefore, a normalizable translational zero mode 
always exists with the mode function $T'(y)$ and the 
massless effective field $\tau_0(x^\mu)$ in 4D, i.e. 
$\tau(x^\mu,y) = T'(y)\tau_0(x^\mu)$.

In the rest of this subsection, we will study mass spectra of 
the scalar field $\Phi$. 
The linearized equation for small fluctuation $\Phi$ is given by 
(we will use the same notation $\Phi$ for the fluctuation): 
\be
\p_M\left(
\beta(T)^2
\p^M\Phi\right) = 0.
\label{eq:leom_Phi}
\ee
First of all, we introduce a canonically normalized field 
$\varphi$ 
\be
\Phi = \frac{\varphi}{\sqrt{2}\beta}. 
\ee
This nonlinear field redefinition transforms 
Eq.~(\ref{eq:leom_Phi}) into 
\be
\left(\square + D^\dagger D
\right) \varphi = 0,
\ee
where we defined
\be
D 
= -\p_y + {\cal Z}'(y),\qquad 
D^\dagger 
= \p_y + {\cal Z}'(y),
\label{eq:D_op}
\ee
with a 1D SQM superpotential
\be
{\cal Z}(y) = \frac{1}{2}\int^y d\lambda\, \frac{d}{d\lambda} \log \beta(T(\lambda))^2 = \frac{1}{2}\log \beta(T(y))^2.
\label{eq:Z_function}
\ee
Note that this is valid for any background solutions since we 
have not used the (anti)BPS equation.
Thus, we have obtained another 1D eigenvalue problem with the 
SQM structure 
\be
D^\dagger D h_n = \mu_n^2 h_n,
\label{eq:SQM_scalar2}
\ee
Unlike the fermionic case, the super partner $D D^\dagger$ is 
absent in the problem.

The solution with zero eigenvalue is unique and is given by 
\be
h_0(y) = e^{\int^y d\lambda\, {\cal Z}'(\lambda)} =  e^{{\cal Z}(y)} =  
\beta(T(y)).
\label{eq:h_zero}
\ee
This is a normalizable physical state whenever  
$\beta(T(y))$ is square integrable. 
We will call the massless scalar boson topological only in 
the limited sense that 
it is stable against small 
changes of parameters in the nonlinear kinetic function $\beta$.
As is clear from the derivation, it is not the NG boson for 
the spontaneously broken rigid symmetry such as translation. 
We observe that the 1D eigenvalue problem for mass spectra of 
scalar field becomes identical to that of fermion
by identifying the function 
${\cal Z}'(y)= d\log\beta(T(y))/dy$ in the
operator $D$ with 
${\cal Y}'(y) = {\cal M}(T(y))$  in the operator ${\cal Q}$
\be
{\cal Y}'(y) = {\cal M}(T(y))\quad \leftrightarrow \quad 
{\cal Z}'(y)= \frac{d\log\beta(T(y))}{dy}.
\ee
We assume that the function ${\cal Z}'(y)$ goes across 
zero as ${\cal M}(y)$ in Fig.~\ref{fig:fermion_zero}.
Namely, the function ${\cal Z}'(y)$ satisfies 
the 
%gap
following condition as in the fermion case in Eq.~(\ref{eq:opp_asymp_value}) 
\be
{\cal Z}'(y=-\infty)\times {\cal Z}'(y=+\infty)<0. 
\label{eq:opp_asymp_value_scalar}
\ee
In the present case of scalar field, we have to choose 
${\cal Z}'(y=+\infty)<0$ 
for $h_0$ to be normalizable.\!\footnote{
A weaker boundary condition is allowed for normalizability. 
The asymptotic value of ${\cal Z}'$ can vanish, for instance 
${\cal Z}'(y)\sim -\alpha/y, \; \alpha>1/2$ for $y\to\infty$, 
instead of a nonvanishing constant ${\cal Z}'(y=\infty)<0$. 
This weaker condition is also valid for ${\cal M}$ in 
Eq.~(\ref{eq:opp_asymp_value}) for fermions.
} 
In the opposite case with ${\cal Z}'(y=+\infty)>0$, there are no 
normalizable massless modes.

We now come to a highlight of this work.
We define the Jackiw-Rebbi-like mechanism for bosons as follows:
{\it When the field-dependent ``coupling'' $\beta(T)$ defined
in Eq.~(\ref{eq:L_scalar}) satisfies the 
%gap 
condition 
%given in Eq.~
(\ref{eq:opp_asymp_value_scalar}), 
a localized massless scalar boson appears and is 
localized around a point where $d\beta(T(y))/dy$ 
vanishes.}
Similarly to the fermion case, 
the massless boson is stable against any continuous changes 
which do not violate the %gap 
condition 
(\ref{eq:opp_asymp_value_scalar})
for $\beta'/\beta$.
In short, the massless scalar field in 
Eq.~(\ref{eq:h_zero}) is a topological edge 
state which is supported by the Jackiw-Rebbi-like mechanism for bosons.

%%%%%%%%%%%%%%%%%%%%%%%%%%%%%%%%%%%

\subsection{Domain wall vectors/scalars
}

In this section we consider (1-form) gauge fields. We consider a 
gauge invariant Lagrangian similar to ${\cal L}_0$ in
Eq.~(\ref{eq:L_scalar}),
\be
{\cal L}_1 = -\beta(T)^2 {\cal F}_{MN}{\cal F}^{MN}.
\label{eq:LG}
\ee
Here, we only consider an Abelian gauge field ${\cal A}_M$ with 
the field strength ${\cal F}_{MN} = \p_M {\cal A}_N - \p_N {\cal A}_M$ 
just for simplicity, but it is straightforward to extend the 
following results to Yang-Mills fields \cite{Arai:2018rwf}.

As was explained in the Introduction, the Lagrangian (\ref{eq:LG}) is a model for the localized gauge 
fields on domain walls in the brane-world-scenario. 
To localize gauge fields on topological defects like domain walls, 
it was recognized that the confining phase is needed in 
the bulk, and a toy model in four spacetime dimensions was 
explicitly proposed \cite{Dvali:1996xe}. 
The field-dependent kinetic term for gauge fields was 
considered together with further explicit toy model in six spacetime 
dimensions \cite{ArkaniHamed:1998rs}, and 
an explicit model has been constructed in 
five spacetime dimensions \cite{Ohta:2010fu}.
Another study for nonsupersymmetric model with/without gravity 
was developed in \cite{Chumbes:2011zt}.
The coefficient $\beta$ in Eq.~(\ref{eq:LG}) 
can be considered as an inverse of the position dependent gauge 
coupling after the scalar field $T$ takes a nontrivial $y$-dependent 
values as the background. 
Bulk with $\beta = 0$ implies infinitely large gauge coupling, 
which is a semiclassical realization of the confining vacuum 
\cite{Kogut:1974sn,Friedberg:1976eg,Friedberg:1977xf,Friedberg:1978sc,
Fukuda:1977wj,Fukuda:2008mz,Fukuda:2009zz}. 
Due to the so-called dual Meisner effect, (chromo)electric 
field cannot invade the bulk, so that massless gauge fields 
are confined inside a finite region (for us it is inside 
the domain wall) where $\beta$ is not zero.

\vspace*{.5cm}
Leaving aside the above qualitative interpretation of the model 
based on a somewhat speculative intuition of confinement in dimensions 
higher than four, we will now focus on the underlying mathematical structure 
of the localization mechanism inherent in the model (\ref{eq:LG}). 
It is very close to the model of topological massless scalar fields 
in Sec.~\ref{eq:DWS}.
Namely, {\it the massless gauge field is supported by the 
Jackiw-Rebbi-like mechanism for bosons}. 
In order to see the relation clearly, 
let us investigate the mass spectrum of the gauge field 
about the domain wall background $T(y)$.
Firstly, we need to fix unphysical gauge degree of freedom. 
The most popular gauge choice is the axial gauge ${\cal A}_y=0$, 
see for example Refs.~\cite{Ohta:2010fu,Chumbes:2011zt}.
However, one should be careful to deal with a possible 
normalizable zero mode in ${\cal A}_y$, since, 
if it exists, it is gauge invariant and cannot be gauged away. 
Therefore, one cannot fully remove ${\cal A}_y(x,y)$ before 
confirming the absence of normalizable zero modes. 
To clarify this point, we have developed a new gauge fixing 
condition recently by adding the following gauge fixing term 
\cite{Okada:2017omx,Arai:2018uoy,Okada:2018von}
\be
{\cal L}_{\rm GF} = - \frac{2}{\xi} \beta(T)^2 
\left[\p_\mu{\cal A}^\mu - \frac{\xi}{\beta(T)^2} 
\p_y\left(\beta(T)^2 {\cal A}_y\right)\right]^2,
\label{eq:GF_A}
\ee
where $\xi$ is an arbitrary gauge fixing parameter. 
We call this the extended $R_\xi$ gauge \cite{Arai:2018uoy}.

To study the mass spectra, let us consider small fluctuations 
$A_M$ around the domain wall background and we define a 
canonically normalized fields in five spacetime dimensions, 
which will make the analogy to the Jackiw-Rebbi mechanism 
most explicit 
\be
{\cal A}_M = \frac{A_M}{2\beta(T)}.
\ee
Then the linearized equations of motion in the generalized 
$R_\xi$ gauge are given by \cite{Arai:2018uoy}:
\begin{align}
&\left[\eta^{\mu\nu}\square - \left(1-\frac{1}{\xi}\right)
\p^\mu\p^\nu + \eta^{\mu\nu}D^\dagger D\right]A_\nu = 0,
\label{eq:gauge_4}\\
&\left(\square + \xi D D^\dagger \right) A_y = 0.
\label{eq:gauge_5}
\end{align}
We again encounter $D$ and $D^\dagger$ defined in Eq.~(\ref{eq:D_op}). 
However, not only $D^\dagger D$ but also $D D^\dagger$ comes 
into play, unlike the case of the scalar field.
Thus, the 1D eigenvalue problem for mass spectra exhibits the 1D SQM 
structure in precise analogy with the Jackiw-Rebbi mechanism 
for fermions 
\be
D^\dagger D h_n = \mu_n^2 h_n,\qquad
D D^\dagger \tilde h_n = \mu_n^2 \tilde h_n.
\label{eq:Sch_A}
\ee
As before, the eigenvalue spectra of $D^\dagger D$ and $DD^\dagger$ 
coincide except for zero eigenvalue. 
We observe that the massive modes of $A_y$ are unphysical, since 
their masses depend on the gauge-fixing parameter $\xi$, and will be 
cancelled by the ghost fields with the same mass. 
However, $n=0$ is special. 
Eq.~(\ref{eq:gauge_5}) shows that the zero mode $A_y^{(n=0)}$ 
of $A_y$ is just a massless scalar field.
The gauge fixing parameter $\xi$ disappears from Eq.~(\ref{eq:gauge_5}), 
so that $A_y$ of $n=0$ is not a gauge-dependent degree 
of freedom. The observation that the normalizable 
zero mode of $A_y$ can be physical scalar field is missed in many 
previous works using ${\cal A}_y = 0$ gauge.
For the zero mode $n=0$ of $A_\mu$, Eq.~(\ref{eq:gauge_4}) reduces to 
the linearized equation for the massless photon in the usual 
covariant gauge.

The mode functions with the zero eigenvalue are explicitly given by
\be
h_0(y) = \beta(T(y))
,\qquad \tilde h_0(y) = \frac{1}{\beta(T(y))}\,.
\label{eq:zero_boson}
\ee
Thus, as in the scalar case, the physical massless gauge field 
appears on the domain wall whenever $\beta(T(y))$ is square integrable. 
This is the case when 
the %gap 
condition Eq.~(\ref{eq:opp_asymp_value_scalar}) 
is satisfied with ${\cal Z}'(y=+\infty)<0$. 
On the other hand, $\tilde h_0(y)$ is not normalizable, as long 
as we consider noncompact space $-\infty<y<\infty$. 
Hence $A_y$ does not supply a physical massless scalar field. 
Up to this point, the final result turns out to be the same as 
that obtained in the axial gauge ${\cal A}_y =0$. 
However, there are two other possibilities.  

The first possibility is that 
the %gap 
condition in Eq.~(\ref{eq:opp_asymp_value_scalar}) 
is satisfied with ${\cal Z}'(y=+\infty)>0$. Then
the physical massless field 
localized on the domain wall is scalar, 
since $\beta(T(y))^{-1}$ 
is square integrable. 
In this case, the massless vector field becomes unphysical 
because it is no longer normalizable. 
Thus, the spin of massless bosons is determined 
by the sign of the asymptotic value of the function ${\cal Z}'(y=+\infty)$ : 
The massless boson is vector if ${\cal Z}'(y=+\infty)<0$ or is scalar 
if ${\cal Z}'(y=+\infty)>0$, 
similarly to the selection of chirality in the case of the 
Jackiw-Rebbi mechanism for fermions. 

Another possibility is to consider compact space such as the circle 
for the extra dimension $y$. 
We will discuss this possibility in 
Sec.~\ref{sec:non-BPS}.

%%%%%%%%%%%%%%%%%%%%%%%%%%%%%%%%%%%

\subsection{Domain wall tensors/vectors}

Let us now consider a two-form field in five dimensions with 
the Lagrangian 
\be
{\cal L}_2 = \beta(T)^2 {\cal H}_{MNL}{\cal H}^{MNL}.
\ee
Here, we consider a two-form field $\theta_{MN} = - \theta_{NM}$ 
with a field strength ${\cal H}_{MNL} = \partial_M \theta_{N L}
+ \partial_L \theta_{MN}+ \partial_N \theta_{L M}$. The above 
Lagrangian is invariant under the gauge transformation 
$\theta_{MN}\to \theta_{MN}+\partial_M \Lambda_N-\partial_N\Lambda_M$, 
where $\Lambda_M$ is an arbitrary $U(1)$ gauge field. 
To fix the gauge and clarify unphysical degrees of freedom, 
we choose to add the following gauge-fixing terms\footnote{
Similar analysis but in the different gauge $\theta_{\mu y} = 0$ 
was done in \cite{Chumbes:2011zt}.
However, it will turn out that this gauge fixing misses the 
possibility of appearance of massless modes in the $\theta_{\mu y}$ 
component.
}
\be
{\cal L}_{\rm GF} = \frac{6}{\xi}\beta(T)^2
\left(\partial_\mu\theta^{\mu\nu}+\frac{\xi}{\beta(T)^2} 
\partial_y\bigl(\beta(T)^2
\theta_{\phantom{\nu}y}^{\nu}\bigr)\right)^2
-\frac{6}{\eta}\beta(T)^2\bigl(\partial_\mu 
\theta_{\phantom{\mu}y}^{\mu}\bigr)^2.
\label{eq:GF_theta}
\ee
Similarly to the generalized $R_\xi$ gauge employed in the 
previous section, these terms are devised in such a way as to 
eliminate the mixing terms between extra-dimensional and 
four-dimensional components. 
Notice that we have two independent 
gauge-fixing parameters, namely $\xi$ and $\eta$.

Let us investigate mass spectra of fluctuation fields
of $\theta_{MN}$ around the domain wall background. 
In terms of the canonically normalized fields
\be
\theta_{\mu\nu} = \frac{h_{\mu\nu}}{\beta(T)}, \hspace{5mm} 
\theta_{\mu y} = \frac{B_\mu}{\sqrt{12}\, \beta(T)}
\ee
the linearized equations of motion read
\begin{align}
& \Bigl[\eta^{\mu\rho}\eta^{\nu\sigma}\square+\eta^{\mu\sigma}\partial^\rho\partial^\nu +\eta^{\nu\rho}\partial^\sigma\partial^\mu+\tfrac{2}{\xi}\eta^{\nu\sigma}\partial^\mu\partial^\rho+\eta^{\mu\rho}\eta^{\nu\sigma}D^{\dagger}D \Bigr]h_{\rho\sigma} = 0, \\
& \Bigl[\eta^{\mu\nu}\square-\bigl(1-\tfrac{1}{\eta}\bigr)\partial^\mu\partial^\nu+\xi\eta^{\mu\nu}D D^{\dagger}\Bigr]B_\nu = 0.
\end{align}
Thus, no new 1D eigenvalue 
problems arise as the differential operators $D$ 
and $D^{\dagger}$ are the same as for scalar (zero-form) and vector (one-form) fields.

Similarly to the vector fields, 
existence of physical massless modes is guaranteed by the %gap 
condition 
in Eq.~(\ref{eq:opp_asymp_value_scalar}). 
Namely, the spin of the physical massless 
bosons is determined by the sign of the asymptotic value of 
the function ${\cal Z}'(y=+\infty)$ : 
Only the tensor field $\theta_{\mu\nu}$ has a zero mode if 
${\cal Z}'(y=+\infty)<0$ since $\beta(T(y))$ is square integrable, whereas only the vector field $\theta_{\mu y}$ 
has a zero mode if ${\cal Z}'(y=+\infty)>0$ since $\beta^{-1}(T(y))$ is square integrable.

Let us consider the case of ${\cal Z}'(y=+\infty)<0$, where we 
have the massless mode $h_{\mu\nu}^{(0)}$. 
From the four-dimensional point of view of effective field theory, 
the massless mode can be understood as a scalar field via a 
duality, 
\be
\partial_\mu h_{\nu\rho}^{(0)}+\partial_\rho h_{\mu\nu}^{(0)}
+\partial_\nu h_{\rho\mu}^{(0)} 
= \varepsilon_{\mu\nu\rho\sigma}\partial^\sigma \phi\,,
\ee
where $\phi$ is a massless scalar. 
On the other hand, the massive states $h_{\mu\nu}^{(n)}$ can be 
interpreted as massive vector fields,
whereas all the massive states 
in the second tower $B_\mu^{(n)}$ are unphysical as their masses 
are proportional to $\xi$.

In contrast, if ${\cal Z}'(y=+\infty)>0$, 
the normalizable zero mode $B_\mu^{(0)}$ now exists. 
It is easy to see that $B_\mu^{(0)}$ acts as a gauge field under 
$y$-independent gauge transformations of $\theta_{MN}$ and, 
therefore, there is a localized $U(1)$ gauge field in the spectrum.

In the case of ${\cal Z}'(y=+\infty)<0$ ($\beta$ being square 
integrable), the spectrum of localized particles for two-form field 
is a massless dual scalar $\phi$ and a tower of massive vector 
fields dual to $h_{\mu\nu}^{(n\neq0)}$. 
This spectrum is identical to the spectrum for one-form field 
($A_y^{(0)}$ and $A_\mu^{(n\neq0)}$) in the case of 
${\cal Z}'(y=+\infty)>0$ ($1/\beta$ being square integrable), 
as shown in the previous section. 

Similarly, if ${\cal Z}'(y=+\infty)>0$
for two-form field, we have the spectrum of a massless gauge 
field $B_\mu^{(0)}$ and a tower of massive vector 
fields dual to $h_{\mu\nu}^{(n\neq0)}$, 
which precisely coincides 
with the spectrum ($A_\mu^{(0)}$ and $A_\mu^{(n\neq0)}$)
for one-form in the case of 
${\cal Z}'(y=+\infty)<0$.

This correspondence can be easily understood via on-shell duality 
between two-forms and one-forms in five dimensions. 
Indeed, if we look at the full equation of motion 
\be
\partial_M\left(\beta^2{\cal H}^{MNL}\right) = 0\,,
\ee  
we can solve it by setting
\be
{\cal H}_{MNL} = \beta^{-2}\bigl(\varepsilon_{MNLPQ}{\cal F}^{PQ}\bigr)\,,
\ee
where ${\cal F}^{PQ} = \partial^P {\cal A}^Q-\partial^Q {\cal A}^P$ and ${\cal A}_M$ is some gauge field. Note that the Bianchi identity
\be
\varepsilon^{MNLPQ}\partial_N {\cal H}_{LPQ} = 0
\ee
translates into the equation of motion for the gauge field, i.e. 
$\partial_M\bigl(\beta^{-2} {\cal F}^{MN}\bigr) = 0$ which is the 
same equation of motion as in the previous section but it comes 
with $\beta^{-2}$ in place of $\beta^2$. 

%%%%%%%%%%%%%%%%%%%%%%%%%%%%%%%%%%%

\section{Simple models}
\label{sec:example}

\subsection{A class of calculable models}
\label{sec:cal}

As we have stressed so far, there are no strong constraints for 
both ${\cal M}(T)$ and $\beta(T)$. 
However, it is extremely convenient to choose a particular form 
in order to gain a calculability even in the case of non-BPS 
background solution. 
One of the simplest example we choose is
\be
{\cal M} (T) = \epsilon_{\rm F} W_{TT}(T),\qquad
\beta (T) = W_T(T)^{\epsilon_{\rm B}}\,,
\label{eq:SUSY}
\ee
where $\epsilon_{\rm B,F}$ is either $+1$ or $-1$.
With the choice of ${\cal M}(T)$, ${\cal L}_{\rm DW} + {\cal L}_{\rm F}$ 
is close to the Wess-Zumino SUSY model in $D=4$. 
However, it is not our intention to stick to genuine supersymmetric 
models in five spacetime dimensions. 
Instead, we only use the model to gain calculability hoping to 
get general qualitative features in a simple and transparent 
manner without being constrained by supersymmetry. 

In the rest of this section, we will focus on the BPS domain 
wall which satisfies $T' = W_T(T)$. 
The case of antiBPS domain wall is straightforward, and 
nonBPS cases will be studied in Sec.~\ref{sec:non-BPS}. 
The translational NG boson is given in Eq.~(\ref{eq:trans_0_BPS}).

The normalizable fermionic zero mode given in 
Eq.~(\ref{eq:fermion_zero_mode}) reads
\be
f_0(y)\big|_{\rm BPS} &=& e^{\epsilon_{\rm F} \int^y d\lambda\, W_{TT}(T(\lambda))}  = W_T(T(y))^{\epsilon_{\rm F}},\\
\tilde f_0(y)\big|_{\rm BPS} &=& e^{-\epsilon_{\rm F} \int^y d\lambda\, W_{TT}(T(\lambda))}  =  W_T(T(y))^{-\epsilon_{\rm F}},
\ee
where we have used the BPS equation.
Thus, when $\epsilon_{\rm F}=+1(-1)$, the left-handed (right-handed) massless fermion appears on the domain wall.
Interestingly, the normalizable 
mode functions for the NG boson (\ref{eq:trans_0_BPS}) coincides with
that of the topological fermion. This is due to the SUSY-like structure in ${\cal L}_{\rm DW} + {\cal L}_{\rm F}$.
Namely, the normalizable bosonic  and fermionic zero mode can be 
regarded as  ``supersymmetric'' partners.

The bosonic solutions with zero eigenvalue in 
Eq.~(\ref{eq:zero_boson}) for the choice of $\beta$ in 
Eq.~(\ref{eq:SUSY}) read
\be
h_0(y)\big|_{\rm BPS} = W_T(T(y))^{\epsilon_{\rm B}},\qquad
\tilde h_0(y)\big|_{\rm BPS} = W_T(T(y))^{-\epsilon_{\rm B}},
\ee
where we have not used the BPS equation. Thus, when $\epsilon_{\rm B} = + 1$, 
there exist a massless scalar $\Phi$, vector $A_\mu$, and 
a tensor $\theta_{\mu\nu}$ gauge field on the domain wall for 
${\cal L}_{0,1,2,}$, respectively. 
On the other hand, when $\epsilon_{\rm B} = -1$, no normalizable 
zero modes exist for ${\cal L}_0$, and  a scalar $A_y$ and vector 
$\theta_{\mu y}$ massless modes appears for ${\cal L}_{1,2}$, 
respectively. 
Although there is no obvious hint of supersymmetry between the 
nonlinear kinetic function in Lagrangians ${\cal L}_{0,1,2,}$ 
and ${\cal L}_{\rm DW}$ or ${\cal L}_{\rm F}$, the mode function 
of the topological bosons turn out to coincide with those of 
the translational NG boson and the topological massless fermion. 
The only link that one can find is the SQM structure common to 
all these fields in the case of the BPS background solution. 
The mass spectra coincide not only for 
the massless mode but also for all the massive Kaluza-Klein states, since
the 1D SQM superpotentials which determine the mass spectra are 
common to all fields for the BPS domain wall, i.e.
\be
{\cal X}(y)\big|_{\rm BPS} = 
{\cal Y}(y)\big|_{\rm BPS} = 
{\cal Z}(y)\big|_{\rm BPS} = \frac{1}{2}\log W_T(T(y))^2.
\ee

%%%%%%%%%%%%%%%%%%%%%%%%%%%%%%%%%%%

\subsection{Sine-Gordon domain wall}

The simplest example is the sine-Gordon model with the superpotential
\be
W(T) = \frac{\Lambda^3}{g^2}\sin \frac{g}{\Lambda}T.
\label{eq:W_sG}
\ee
The BPS domain wall solutions satisfying $T' = W_T$ are given by
\be
T(y) = \frac{\Lambda}{g} \left(2 \arctan e^{\Lambda y} - \frac{\pi}{2} + 2n\pi\right)
\to \left\{
\begin{array}{ccc}
\left(- \frac{\pi}{2} + 2n\pi\right)\frac{\Lambda}{g} && y \to -\infty\\
\left(\frac{\pi}{2} + 2n\pi\right) \frac{\Lambda}{g} && y \to \infty
\end{array}
\right.\,.
\ee
For these solutions, we have
\be 
W_T(T(y)) = \frac{\Lambda^2}{g}{\rm sech}\, \Lambda y,\qquad
W_{TT}(T(y)) = - \Lambda \tanh \Lambda y.
\ee
There are another set of the BPS solutions given by
\be
T(y) = \frac{\Lambda}{g} \left(2 \arctan e^{-\Lambda y} + \frac{\pi}{2} + 2n\pi\right)
\to \left\{
\begin{array}{ccc}
\left(\frac{3\pi}{2} + 2n\pi\right)\frac{\Lambda}{g} && y \to -\infty\\
\left(\frac{\pi}{2} + 2n\pi\right) \frac{\Lambda}{g} && y \to \infty
\end{array}
\right.\,.
\ee
For these solutions, we have
\be 
W_T(T(y)) = -\frac{\Lambda^2}{g}{\rm sech}\, \Lambda y,\qquad
W_{TT}(T(y)) = - \Lambda \tanh \Lambda y.
\ee
The fact that $W_{TT}(T(y))$ goes across 0 once ensures presence of  the topological massless states.

Since the background is BPS, all the 1D SQM superpotentials agree. Therefore, the mass spectra are determined only 
by $W_{TT}$ in the operator $Q = -\p_y + W_{TT}(T(y))$.
The corresponding SQM Hamiltonians for both BPS solutions are  given by
\be
Q^\dagger Q = - \p_y^2 + \Lambda^2\left(2\tanh^2\Lambda y -1\right),\qquad
Q Q^\dagger = - \p_y^2 + \Lambda^2.
\ee
We have ${\cal Q}^\dagger {\cal Q} = Q^\dagger Q$ and ${\cal Q}{\cal Q}^\dagger = QQ^\dagger$ for $\epsilon_{\rm F} = +1$,
while ${\cal Q}^\dagger {\cal Q} = QQ^\dagger$ and ${\cal Q}{\cal Q}^\dagger = Q^\dagger Q$ for $\epsilon_{\rm F} = -1$.
Similarly, we also have
$D^\dagger D = Q^\dagger Q$ and $DD^\dagger = QQ^\dagger$ for $\epsilon_{\rm B} = +1$,
while $D^\dagger D = QQ^\dagger$ and $DD^\dagger = Q^\dagger Q$ for $\epsilon_{\rm B} = -1$.
Therefore, there exist a unique discrete bound state, which is 
nothing but the normalizable zero mode for 
$\epsilon_{\rm F} = \epsilon_{\rm B} = +1$, 
\be
g_0 = f_0 = h_0 \propto W_T \propto {\rm sech}\Lambda y.
\ee
For the other choice of $\epsilon_{\rm F,B}$, one should replace $f_0$ ($h_0$) by $\tilde f_0$ ($\tilde h_0$).
There are no other discrete states both in the  $Q^\dagger Q$ and 
$QQ^\dagger$ sectors. All the massive modes are continuum
 states (scattering in the bulk) given as
\be
f_k &=& Q^\dagger  e^{iky}  = \left(ik -\Lambda \tanh\Lambda y\right) e^{iky},\\
\tilde f_k &=& e^{iky},
\ee 
with the mass square
\be
m(k)^2 = k^2 + \Lambda^2.
\ee

\subsection{$T^4$ domain wall}

Our second example is the $T^4$ domain wall in the model with cubic super potential
\be
W(T) = \frac{\Lambda^2}{g}T-\frac{g}{3}T^3.
\ee
The BPS domain wall solution is given by
\be
T(y) = \frac{\Lambda}{g} \tanh \Lambda y.
\ee
For this background, we have
\be
W_T(T(y)) = \frac{\Lambda^2}{g} {\rm sech}^2 \Lambda y,\qquad
W_{TT}(T(y)) = - 2 \Lambda \tanh \Lambda y.
\ee
The factor 2 appears compared to the sine-Gordon model. The factor 2 corresponds to the
number of the localized modes as we will see below.

As before, it is enough to investigate ${\cal Q}^\dagger {\cal Q}$ and ${\cal Q}{\cal Q}^\dagger$ because the background is BPS.
We have
\be
{\cal Q}^\dagger {\cal Q} = - \p_y^2 + 2 \Lambda^2\left(3\tanh^2\Lambda y - 1\right),\qquad
{\cal Q}{\cal Q}^\dagger = -\p_y^2 + 2 \Lambda^2 \left(\tanh^2 \Lambda y + 1\right).
\ee
There is unique normalizable zero mode in the 
${\cal Q}^\dagger {\cal Q}$ sector
\be
f_0 \propto W_T \propto {\rm sech}^2 \Lambda y.
\ee
Also there exist a massive discrete state
\be
f_1 &\propto& {\cal Q}^\dagger {\rm sech}\Lambda \propto \tanh\Lambda y~ {\rm sech} \Lambda y,\\
\tilde f_1 &\propto& {\rm sech}\Lambda y.
\ee
All the other states are  continuum states (scattering in the bulk).

%%%%%%%%%%%%%%%%%%%%%%%%%%%%%%%%%%%

\section{Non-BPS domain walls in compact extra dimension}
\label{sec:non-BPS}

\subsection{Quasi solvable example}

So far, we have only considered models with flat non-compact 
extra dimension.
In this section we will study physical spectra about the domain 
walls in compact extra dimension.
For simplicity, we consider the extra dimension to be $S^1$ with a 
radius $R$. 
Unlike the non compact case, all the mode functions are, of 
course, normalizable if they are regular.
Since the profile function $T(y)$ should be periodic, the 
background solution has to be non-BPS which 
includes both BPS and antiBPS domain walls.

\begin{figure}[t]
\begin{center}
\includegraphics[width=15cm]{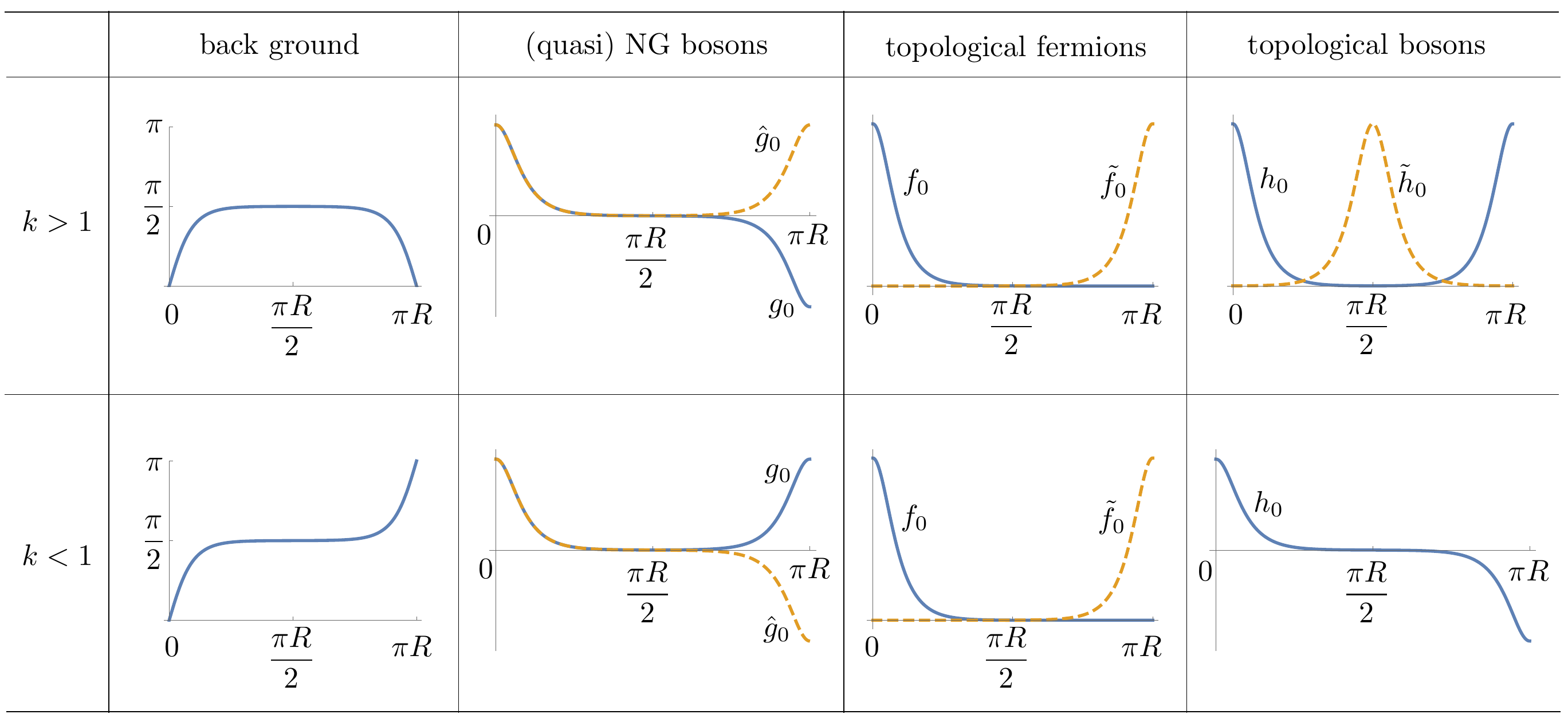}
\caption{Non-BPS domain wall solutions with the BPS and antiBPS walls at $y= 0$ and $\pi R$ in the sine-Gordon model. 
The left-most column shows the background configuration $gT/\Lambda$,
and the other three columns show mode functions of the small fluctuations for the case of $\epsilon_{\rm F,B} = +1$. In the figures,
only the half period is shown. The case of $\epsilon_{\rm F,B} = -1$ can be obtained by exchanging $(f_0,h_0)$ by $(\tilde f_0,\tilde h_0)$.}
\label{fig:nonBPS}
\end{center}
\end{figure}

To be concrete, let us again consider 
the sine-Gordon model with the superpotential given in 
Eq.~(\ref{eq:W_sG}). 
A non-BPS solution with multiple domain walls 
is known \cite{Maru:2001gf} as
\be
T(y) = \frac{\Lambda}{g} {\rm am}\left(\frac{\Lambda}{k}y,k\right),
\ee
where ${\rm am}(x,k)$ denotes the Jacobi amplitude function 
with a real parameter $k$.
Since $T$ can be regarded as an angular variable with periodicity 
$2\pi\Lambda/g$, we can identify the compactification radius $R$ as 
\be
2\pi R = \frac{4k\,K(k)}{\Lambda}, 
\ee
where $K(k)$ is the complete elliptic integral of the first kind.
The solution has BPS and antiBPS domain walls alternatively 
sitting at anti-podal points of $S^1$. 
Namely, the BPS domain wall sits at the origin $y=0$ whereas 
the antiBPS domain wall sits at $y=\pi R$.
The background solutions with $k<1$ and $k>1$ are qualitatively 
quite different ($k=1$ corresponds to either BPS or antiBPS), 
see Fig.~\ref{fig:nonBPS}.
$|gT/\Lambda|$ never goes across $\pi/2$ for the $k>1$ case, 
whereas it monotonically increases (decreases) 
for the $k<1$ case.

Since the above solution is non-BPS, the (anti)BPS equation $T' = \pm W_T$ is not satisfied. Therefore, mass spectra of the
translational NG bosons, the topological fermions, and the topological bosons split.
Let us start with the fluctuation of $T$. Several light modes are explicitly known as
\be
g_0 &\propto& {\rm dn}\left(\frac{\Lambda y}{k},k\right),\quad m_0^2 = 0,\\
\hat g_0 &\propto& {\rm cn}\left(\frac{\Lambda y}{k},k\right),\quad \hat m_0^2 = \frac{1-k^2}{k^2}\Lambda^2,\\
g_2 &\propto& {\rm sn}\left(\frac{\Lambda y}{k},k\right),\quad m_2^2 = \frac{\Lambda^2}{k^2}.
\ee
Note that $g_0 = T'$ is a genuine translational Nambu-Goldstone mode which is exactly massless.
On the other hand, $\hat g_0$ is quasi Nambu-Goldstone mode which corresponds
to the relative distance (so-called radion). It is tachyonic for $k>1$ while it is massive for $k<1$.
The reason why the quasi zero mode is lifted is that unlike for $g_0$ there is no symmetric reasoning for relative distance moduli.
One can also say that the lifting proves that the 
translational zero modes (genuine translational NG and
relative distance moduli) are not topologically protected.
If they were topological, both $g_0$ and $\hat g_0$ would have remained as massless.
These mode functions are depicted in the 2nd column from the left 
of Fig.~\ref{fig:nonBPS}.

Next, let us see the fermions. 
We chose the coupling function ${\cal M}(T)$ for fermions as 
\be
{\cal M}(T) = \epsilon_{\rm F} W_{TT}(T) .
\ee
Then, normalizable zero modes can be explicitly found as
\be
f_0 &\propto& e^{\epsilon_{\rm F} \int^y d\lambda\, W_{TT}(T(\lambda))} 
= \left[{\rm dn}\left(\frac{\Lambda y}{k},k\right) -  k \,{\rm cn}\left(\frac{\Lambda y}{ k},k\right)\right]^{-\epsilon_{\rm F} },\\
\tilde f_0 &\propto& e^{-\epsilon_{\rm F} \int^y d\lambda\, W_{TT}(T(\lambda))} 
= \left[{\rm dn}\left(\frac{\Lambda y}{ k},k\right) -  k \,{\rm cn}\left(\frac{\Lambda y}{ k},k\right)\right]^{\epsilon_{\rm F} }.
\ee
As is well known, $f_0$ is localized around the BPS domain wall at $y=0$ while $\tilde f_0$ is around
the antiBPS domain wall at $y= \pi R$ for $\epsilon_{\rm F} = +1$, see the third column from left of Fig.~\ref{fig:nonBPS}. 
(The mode functions of zero modes are exchanged for $\epsilon_{\rm F} = -1$.)
They are normalizable since the extra dimension is compact.
Note that unlike the translational NG bosons, both $f_0$ and $\tilde f_0$ remain as genuine massless modes 
since they are topological.

Finally, let us see the gauge bosons for the case
\be
\beta(T) = W_{T}(T)^{\epsilon_{\rm B}}.
\ee
We find the exact normalizable 
zero modes for the topological bosons as
\be
h_0 &\propto& \beta = W_T^{\epsilon_{\rm B}} = {\rm cn}\left(\frac{\Lambda y}{ k},k\right)^{\epsilon_{\rm B}},\\
\tilde h_0 &\propto& \beta^{-1} = W_T^{-\epsilon_{\rm B}} 
= {\rm cn}\left(\frac{\Lambda y}{ k},k\right)^{-\epsilon_{\rm B}}.
\ee
When $k>1$, ${\rm cn}(x,k)$ never goes across 0. 
Therefore, both $h_0$ and $\tilde h_0$ are normalizable. 
The mode function $h_0$ for the zero mode of $A_\mu$ is localized at the 
domain walls at $y=0$ and $\pi R$ while $\tilde h_0$ for $A_y$ 
is localized between them when $\epsilon_{\rm B} =+1$.
If $\epsilon_{\rm B} = -1$, the localized positions of $h_0$ 
and $\tilde h_0$ are exchanged.
When $k<1$, ${\rm cn}(x,k)$ goes across 0. 
Therefore $\tilde h_0$ ($h_0$) is singular and non-normalizable 
for $\epsilon_{\rm B} =+1$ ($\epsilon_{\rm B} = -1$).
We show $h_0$ and $\tilde h_0$ for $\epsilon_{\rm B} = +1$ in the 
right-most column of Fig.~\ref{fig:nonBPS}.

\subsection{Phenomenological implications}

\begin{figure}[t]
\begin{center}
\includegraphics[width=12cm]{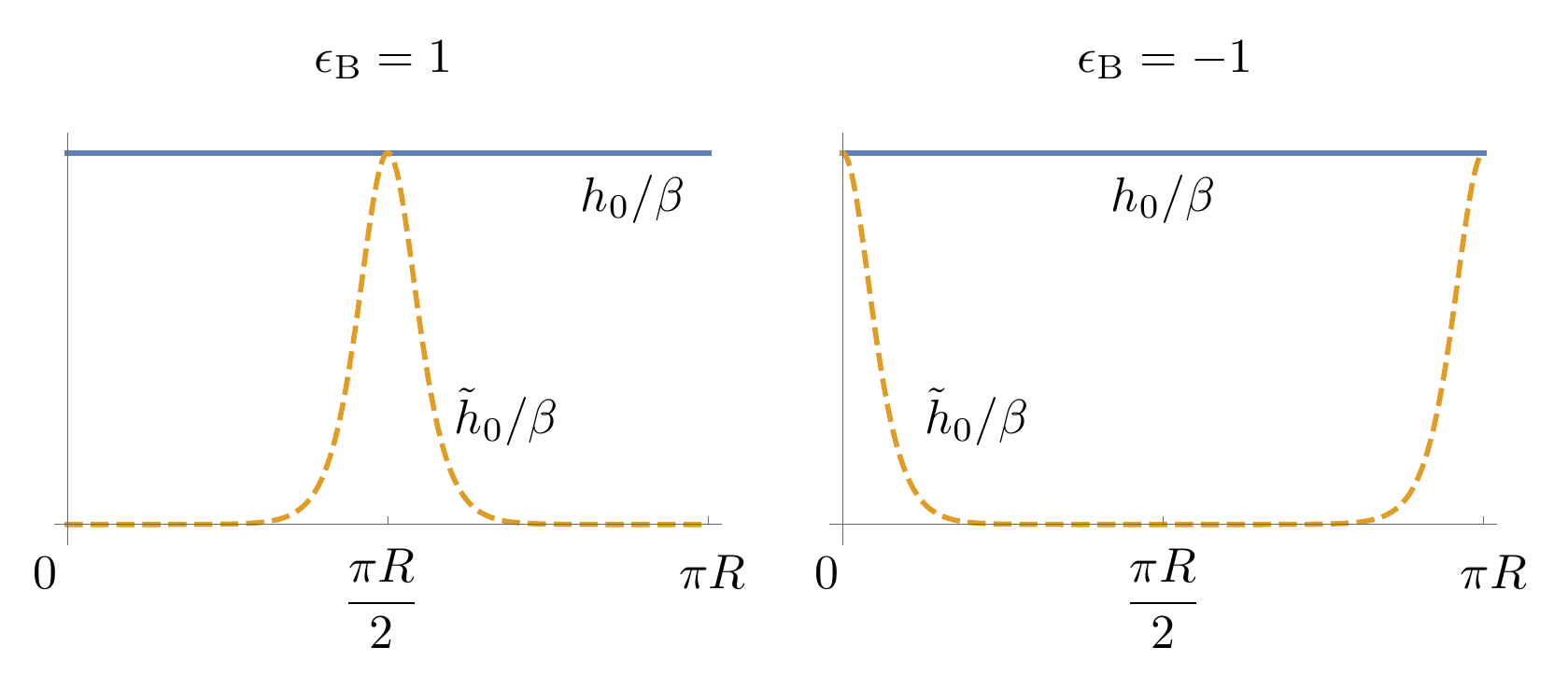}
\caption{
The mode functions of the topological bosons for the non-canonical fields $\Phi,{\cal A}_M, \theta_{MN}$ around
the non-BPS domain wall solution.
The case of $\epsilon_{B} = 1(-1)$ is shown in the left (right) panel.}
\label{fig:nonBPS2}
\end{center}
\end{figure}

As is shown in Fig.~\ref{fig:nonBPS}, the localization positions 
of the topological fermions and topological bosons are 
sharply different.  Interestingly, $h_0$ ($\tilde h_0$) for 
$\epsilon_{\rm B} = +1$ ($\epsilon_{\rm B} =-1$) have non-zero 
support around both the BPS and antiBPS domain walls. 
This leads to several interesting consequences.
Before going to explain this, however, one should be careful about the mode functions: $h_0$ and 
$\tilde h_0$ are the mode functions of the redefined fields 
$\varphi$, $A_\mu$, $A_y$, $h_{\mu\nu}$ and $B_\mu$. 
The mode functions for the original fields $\Phi$, ${\cal A}_M$ 
and $\theta_{MN}$ are those divided by $\beta$, see Fig.~\ref{fig:nonBPS2}.
\be
h_0 \to \frac{h_0}{2\beta} = \text{const.}\,,\qquad
\tilde h_0 \to \frac{\tilde h_0}{2\beta} \propto \beta^{-2} = W_T^{-2\epsilon_{\rm B}}\,.
\ee

In the following, we choose the background solution with $k>1$ 
which is not afflicted by the problem like non-normalizability 
of mode functions. For phenomenology in the brane-world scenario, let us concentrate 
on the (1-form) gauge field in the following.
Suppose that the fermion is charged under the $U(1)$ gauge symmetry 
with unit charge. The covariant derivative is given by 
${\cal D}_M \Psi = \left(\p_M + i {\cal A}_M\right)\Psi$. 
We find the gauge 
interactions of massless fermions as 
\be
\int dy\, \bar\Psi \Gamma^\mu {\cal D}_\mu \Psi  &\ni&
\int dy \biggl(f_0^2\bar \psi_L^{(0)}\gamma^\mu 
\left(\p_\mu + i \frac{h_0}{2\beta}A_\mu^{(0)}\right)\psi_L^{(0)} +
\tilde f_0^2  \bar\psi_R^{(0)}\gamma^\mu 
\left(\p_\mu + i \frac{h_0}{2\beta}A_\mu^{(0)}\right)\psi_R^{(0)}\biggr) \nonumber\\
&=& \bar \psi_L^{(0)}\gamma^\mu \left(\p_\mu + i e_4 A_\mu^{(0)}\right)
\psi_L^{(0)}+ \bar\psi_R^{(0)}\gamma^\mu 
\left(\p_\mu + i e_4 A_\mu^{(0)}\right)\psi_R^{(0)},
\ee
where we have used the fact that $h_0$ is proportional to $\beta$ 
as $h_0 = 2 e_4 \beta$ with
\be
e_4^{-2} = 4 \int_0^{2\pi R} dy\,  \beta^2.
\ee

It is important to notice that the effective gauge coupling 
$e_4$ is universal. It is also independent of the fermion mode functions.
Hence, the low energy effective theory is a vector-like gauge 
theory such as QED or QCD in which the left and right handed
fermions are coupled with the gauge field with the same strength. 
In order to have a chiral gauge theory like the Standard Model in 
our framework, we have to consider the infinitely separated limit 
$R=\infty$ ($k=1$). 
This situation is in accord with the usual notion of domain wall 
fermion in lattice gauge theories.

In contrast to the gauge interactions in four-dimensions, we have an interesting non-universality 
for the coupling of massless scalar coming from $A_y$. 
The induced Yukawa-type coupling of the scalar $A_y^{(0)}$ is given as 
\be
\int dy\, \bar\Psi \Gamma^y {\cal D}_y \Psi  &\ni&
- g_L A_y^{(0)}\bar \psi_L^{(0)}\gamma_5  \psi_L^{(0)} -
g_R A_y^{(0)} \bar\psi_R^{(0)}\gamma_5  \psi_R^{(0)}\,.
\ee
where we used the fact that $\tilde h_0 = 2 \tilde e_4 \beta^{-1}$ as
\be
\tilde e_4^{-2} = 4 \int_0^{2\pi R} dy\,  \beta^{-2}.
\ee
and  defined
\be
g_L &\equiv& \int^{2\pi R}_0 dy\, f_0^2\frac{\tilde h_0}{2\beta} = \int_0^{2\pi R} dy\, \frac{\tilde e_4 f_0^2}{\beta^2},\\
g_R &\equiv& \int^{2\pi R}_0 dy\, \tilde f_0^2\frac{\tilde h_0}{2\beta} = \int_0^{2\pi R} dy\, \frac{\tilde e_4 \tilde f_0^2}{\beta^2}.
\ee
Now, we find that $g_{L(R)}$ plays a role of effective 
Yukawa coupling for scalar field $A_y^{(0)}$. 
 Firstly, since 
$f_0$ ($\tilde f_0$) and $\tilde h_0$ are separately localized 
at different positions as shown in Figs.~\ref{fig:nonBPS},\ref{fig:nonBPS2}
for $\epsilon_{\rm B} = 1$, the 
overlap integrals for $g_{L(R)}$ are exponentially small. 
This can help to explain smallness of the Yukawa couplings for 
the first and second generation of quarks and leptons 
\cite{ArkaniHamed:1999dc}.
Secondly, the scalar field $A_y^{(0)}$ can play a role of the 
Higgs field \cite{Hosotani:1983xw,Hosotani:1983vn}. 
If $A_y^{(0)}$ enjoys a non-zero 
vacuum expectation value (VEV), it immediately means the fermions 
get masses. Since the Higgs field is originated as the 
extra-dimensional gauge field, it is natural to expect that 
quadratic divergences are suppressed thanks to the gauge symmetry 
in the original five-dimensional Lagrangian as advocated by the 
gauge-Higgs unification scenario \cite{Hatanaka:1998yp}. 
In order to verify if $A_y^{(0)}$ actually gets non-zero VEV, one 
must examine an effective potential due to quantum corrections 
such as fermion loop correction. 
We hope to report it in a separate work. 

The results in this section are obtained by using 
a very special simplified model in order to be able to compute 
mode functions and other quantities in a closed form. 
However, we wish to stress that all the qualitative features 
should be valid even if we choose more general functions for 
the coupling functions such as 
${\cal M}(T)$ and $\beta(T)$. 
We only need to use a numerical method to obtain various quantities 
in the general setting. 

%%%%%%%%%%%%%%%%%%%%%%%%%%%%%%%%%%%

\section{Concluding remarks}

Fermionic topological edge (surface) states are well known in 
a vast area of modern physics from high energy physics
to condensed matter physics.  
These fermionic topological states on domain walls are 
robust and are ensured by the Jackiw-Rebbi 
mechanism \cite{Jackiw:1975fn}. 
In this paper, we showed that bosonic topological edge states also 
appear on the domain wall by a quite similar mechanism which we 
call the Jackiw-Rebbi-like mechanism for bosons.
We explicitly showed that it universally 
works for scalar (0-form), vector (1-form), and tensor (2-form) bosonic fields.
They are topological, since their presence only relies 
on boundary condition. 
For localization of vector fields, it has been argued 
that confinement phenomenon is necessary \cite{Dvali:1996bg, 
ArkaniHamed:1998rs, Ohta:2010fu}. 
But it is difficult to show the confinement mechanism especially 
in higher-dimensional field theory. 
On the contrary, the result of this work offers another explanation 
related to topology.
One of the advantages is that it can be applied not only for 
vector but also scalar and antisymmetric tensor fields,
and we can be sure that it works in any spacetime dimensions.  

An interesting feature of the Jackiw-Rebbi(-like) mechanism is that for 
fermions, the domain wall in five dimensions 
selects four-dimensional chirality.
On the other hand, for four-dimensional bosons it selects spin.
For vector (tensor) fields, it selects between four-dimensional 
vector or scalar (tensor or vector).
This can only be seen with the appropriate gauge-fixing terms 
in Eqs.~(\ref{eq:GF_A}) and (\ref{eq:GF_theta}).

We also gave explicit models in Sec.~\ref{sec:example} which 
are useful to see general qualitative features in a simple and 
transparent manner.
Furthermore, we studied massless particles around the non-BPS 
background with a pair of a wall and anti-wall in 
compact extra dimension in Sec.~\ref{sec:non-BPS}.
There, we manifestly showed that the translational zero modes, topological fermionic edge modes,
and topological bosonic edge modes have all different mode functions as is shown in Fig.~\ref{fig:nonBPS}.
We also pointed out possible phenomenological uses of our results. The universality of gauge charges is automatically
satisfied, large hierarchy problem of fermion masses of the Standard Model would naturally be resolved,
and $A_y$ would play a role of the Higgs field as in usual gauge Higgs unification models.

There are several interesting directions for further studies.
In this paper we restricted ourselves in five spacetime dimensions
just for ease of presentation. If we go to higher dimensions than five, 
higher antisymmetric tensor (form) fields can appear. 
We should examine how the selection rules by 
the domain wall is generalized. 
We can also consider other solitons like vortex and monopole 
whose co-dimensions are higher than one. As is the case of domain wall,
localization of topological fermions are well known. 
We will study whether it is true for bosons or not.
On the other hand, it is also very interesting to go to lower dimensions.
If our bosonic topological states are found in a real material, 
it is an indirect proof of 
localization of all the Standard Model particles on a domain wall. 
Apart from the brane-world perspective, 
it might be interesting for revealing new properties of topological matters.
The domain wall fermions are known to be important in lattice QCD, so we also wonder if the topological 
localization mechanism of bosons plays some role for improving computer simulations of lattice QCD.

%%%%%%%%%%   ACKNOWLEDGMENTS   %%%%%%%%%%

\section*{Acknowledgements}
M.\ E. thanks to Hidenori Fukaya and Yu Hamada for discussions.
M.\ E. also thanks the Yukawa Institute for Theoretical Physics 
at Kyoto University. 
Discussions during the YITP workshop YITP-W-18-05 on ``Progress 
in Particles Physics 2018" were useful to complete this work.
This work is supported in part 
by the Japan Society for the Promotion of Science (JSPS) 
Grant-in-Aid for Scientific Research (KAKENHI) Grant Numbers 
No. 16H03984 (M.\ E.), No. 19K03839 (M.\ E.), and 18H01217 (N.\ S.). 
This work is also supported in part by the Ministry of Education, Culture, 
Sports, Science, and Technology (MEXT)-Supported Program for the 
Strategic Research Foundation at Private Universities ``Topological 
Science" (Grant No. S1511006) (N.\ S.), and
also by MEXT KAKENHI Grant-in-Aid for Scientific Research on Innovative Areas  
``Discrete Geometric Analysis for Materials Design" 
No.~JP17H06462 (M.\ E.) from the MEXT of Japan. 
This work was also supported by the Albert Einstein Centre for 
Gravitation and Astrophysics financed by the Czech Science Agency 
Grant No. 14-37086G (F.\ B.) and by the program of Czech Ministry 
of Education Youth
and Sports INTEREXCELLENCE Grant number LTT17018 (F. B.).
F.\ B.\ was an international research fellow of the Japan Society 
for the Promotion of Science, and was supported by Grant-in-Aid 
for JSPS Fellows, Grant Number 26004750.

\bibliographystyle{jhep}

%\bibliography{references}

\end{document}